# Non-scanning large-area Raman imaging for *ex vivo* / *in vivo* skin cancer discrimination

**Elmar Schmälzlin**[a*], **Benito Moralejo**[a], **Ingo Gersonde**[a, b], **Johannes Schleusener**[c], **Maxim E. Darvin**[c], **Gisela Thiede**[c], **Martin M. Roth**[a, d]

[a]Leibniz Institute for Astrophysics Potsdam (AIP), innoFSPEC Potsdam, An der Sternwarte 16, 14482 Potsdam, Germany
[b]innoFSPEC Potsdam, University of Potsdam - Physical Chemistry, Am Mühlenberg 3, 14476 Potsdam, Germany
[c]Charité – Universitätsmedizin Berlin, corporate member of Freie Universität Berlin, Humboldt-Universität zu Berlin, and Berlin Institute of Health, Department of Dermatology, Venerology and Allergology, Center of Experimental and Applied Cutaneous Physiology, Charitéplatz 1, 10117 Berlin, Germany
[d]University of Potsdam, Institute of Physics and Astronomy, Karl-Liebknecht-Str. 24-25, 14476 Potsdam, Germany

**Abstract**. Imaging Raman spectroscopy can be used to identify cancerous tissue. Traditionally, a step-by-step scanning of the sample is applied to generate a Raman image, which, however, is too slow for the routine examination of patients. By transferring the technique of integral field spectroscopy (IFS) from astronomy to Raman imaging, it became possible to record entire Raman images quickly within one single exposure without the need of a tedious scanning procedure. In this work, an IFS-based Raman imaging setup is presented that is capable to measure skin *ex vivo* or *in vivo*. It is demonstrated how Raman images of healthy and cancerous skin biopsies were recorded and analyzed.

**\*** Corresponding Author: E-mail: eschmaelzlin@aip.de

## 1. Introduction

Imaging Raman spectroscopy is a powerful tool to identify chemicals and their distribution. When monochromatic light impinges on molecules, fractions of the scattered light are wavelength shifted related to the molecular vibration states. Thus, Raman spectra are fingerprints that allow a contact- and label-free identification of chemical structures. In contrast to IR absorption spectroscopy that also measures vibration transitions, Raman spectroscopy works in an aqueous environment, which makes this method promising for biological analysis [1, 2, 3], especially in the field of medical diagnostics for an identification of cancerous tissue [4, 5, 6, 7, 8]. In surgical cancer treatment, the determination of resection margins is a much discussed topic [9, 10, 11]. Too spacious tissue removal stresses the patient, while too tight margins reduce the chances of recovery. Usually, a biopsy is taken and examined *ex vivo* by a pathologist. Based on the results the surgeon later makes use of his experience to determine the border between cancerous and healthy tissue. To replace this time consuming two-stage approach and to improve the accuracy, there is a demand for spectroscopic methods that allow a spatially resolved cancer detection *in situ*. Meanwhile, there are medical Raman microscopes available that allow clinical examinations of skin areas on patients directly [12]. However, commercial Raman microscopic systems for *in vivo* skin cancer detection are still single channel systems, i.e. to receive a Raman image from an area of skin, a time-consuming step-by-step scanning process is necessary [13]. Raman intensities are very low (only a fraction in the range of $1 \times 10^{-7}$ of the scattered light is due to Raman scattering) and in addition often superposed by fluorescence. Thus, even for only a few hundred pixels the measurement time typically adds up to many minutes or even hours, which is far too long to perform routine examinations on patients *in vivo*.

To reduce acquisition times, various methods for parallel data collection are described [14]. Of particular interest are full-throughput snapshot techniques, also called "multichannel spectroscopy", "3D spectroscopy" or "integral field spectroscopy" (IFS). These techniques work without any serial scanning procedures and do not sacrifice light fundamentally during the recording procedure. IFS has been developed in astronomy more than three decades ago [15] to save scarce and expensive observation time at observatories. It is based on slicing a two-dimensional image into single strip-like segments and stringing them together to one long row in front of a long-slit spectrograph. This can be e.g. using a mirror stack image slicer or with a fiber bundle converter: At the sample side, the fibers of the bundle are arranged as a two-dimensional matrix. In front of the spectrograph's input, the fiber front surfaces are arranged as a straight line, lying side by side in a V-groove holder. After passing the collimator and camera optics and the dispersive element of the spectrograph, the light signals emerging from each fiber generate a family of individual spectra on a large-area detector. A data reduction software evaluates the raw signal, applies calibrations, and finally provides a data cube containing the entire spectral and spatial information. A review of IFS in astronomy is given in [16]. High-end IFS spectrographs are installed in the MUSE (Multi-Unit Spectroscopic Explorer) system [17] at the Very Large Telescope observatory in Paranal, Chile, since spring 2014. MUSE consists of 24 connected spectrograph modules and is capable of acquiring a total number of 90,000 spatial elements (also called spaxels) within one single exposure. From every single spaxel the entire spectrum from 465 to 930 nm is recorded at a spectral resolution of 0.22 nm in a total of 4300 spectral bins.

To a certain degree, imaging *in vivo* Raman spectroscopy and astronomical observations face similar challenges, namely to detect efficiently faint signals in the presence of bright background light. This notion was the motivation to update a spectrograph based on a MUSE design for use in medical Raman spectroscopy. A fiber bundle-based optical setup was realized to record Raman images of 1 cm$^2$ areas of skin, which matches common sizes of lesions suspected to be cancerous. The objective of the project was to validate the concept of a future instrument that would allow dermatologists to promptly recognize *in vivo* the borders of the cancerous tissue without the need of a time-consuming scanning procedure. The general capability of IFS for generating large-area Raman images was presented previously [18]. Similar Raman setups are described in [19, 20, 21]. Here, we present a discussion and a characterization of a setup using an astronomy spectrograph with regard to Raman imaging of human skin *ex vivo* and *in vivo*. Eventually, comparison measurements of healthy and cancerous human skin samples *in vivo* and *ex vivo* were performed.

## 2. Spectrograph and Optical Setup

Figure 1 shows a scheme of the experimental setup. To record Raman images, an image acquisition head based on two microlens arrays (MLA) was realized. Figure 15 left shows the top of the image acquisition head with a skin sample and the upper MLA underneath. A detailed description of the Raman acquisition optics was given previously [18]. In brief: For excitation, a 784.5 nm diode laser with a tunable output power up to 500 mW was used. Its fiber optic output was connected to a square core fiber with 600 μm core side length. Light emerging from square core fibers shows a top-hat intensity distribution, which is favorable with regard to a preferably homogeneous illumination [22]. The excitation light passes a collimation lens, a clean-up filter to remove Raman and fluorescence background generated within the fiber and spontaneous emission of the laser, and is finally guided to the sample by a 45° dichroic mirror. An MLA in front of the sample generates 20 × 20 excitation spots on the sample with a 0.5 mm pitch, i.e. a square image with 1 cm side length and a sampling of 400 pixels can be recorded. In the opposite direction, the MLA collects the Raman signal from the sample. A pair of relay lenses guides the Raman signal to a further MLA that couples the signal into the fibers of the fiber bundle. A 785 nm notch filter and a long pass (LP) filter remove the Rayleigh signal.

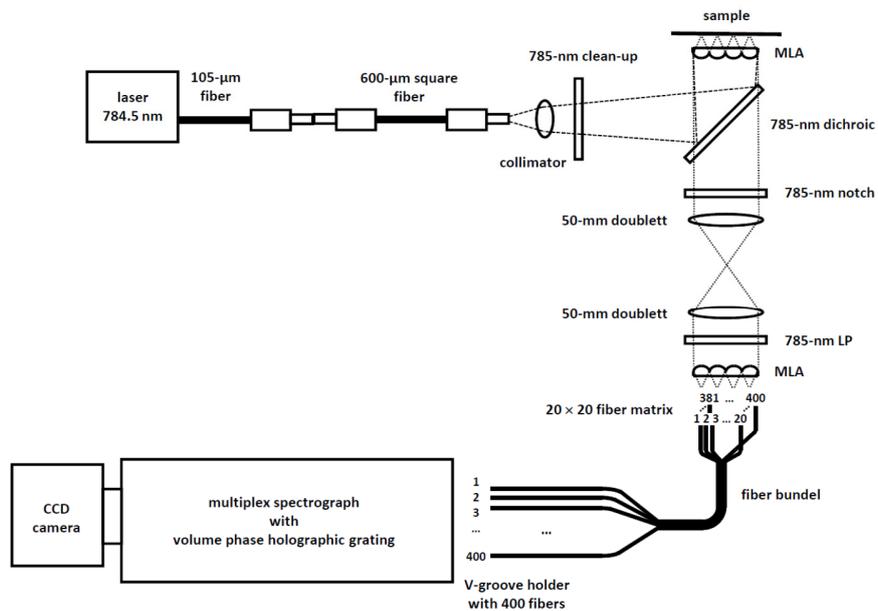

**Fig. 1** Setup for generating Raman images from 1 cm2 skin patches without scanning.
See explanation in the text.

The fiber bundle consists of 400 fibers (114/125/155 VIS/IR, $NA$ = 0.22, Heracle, Germany). On the sample side the fibers are arranged within a square plate containing 20 × 20 microholes at center distances of 0.5 mm (Fig. 2). On the spectrograph input, the fibers are arranged side by side within a V-groove holder forming a pseudo slit at a length of 118 mm. There are 421 V-grooves with a pitch of approximately 0.29 mm. The fibers of the bundle are arranged in groups of twenty. Additional V-grooves between the groups are occupied with fibers that form a fan-out cable that is intended for other applications (calibration, tests). Due to the gaps, the groups of twenty are easily distinguishable in the raw data, thus facilitating a quick inspection by eye before starting the data reduction process.

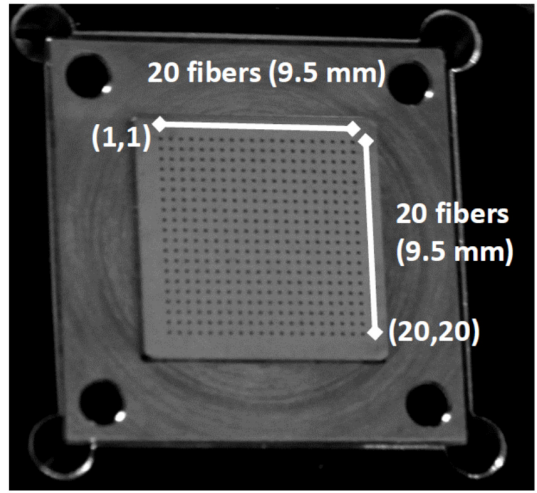

**Fig. 2** Metal housing of the 20 × 20 fiber matrix. Every fiber front surface corresponds to one pixel of the Raman image.

A detailed description of the spectrograph was given previously [23]. The array of optical fibers is attached to a first plane-concave silica lens. The plane side of the lens was initially intended to use index matching gel to minimize coupling losses. However, it was far more advantageous to allow an easy back and forward sliding of the fiber holder. Pushed back, the fibers could be illuminated with a high power white light LED array. On the basis of the resulting light spots, the optical components of the head, especially the positions of the MLAs, could be adjusted. During the alignment procedure, the 785 nm long pass filter was flipped. To allow a routinely check of the alignment, the index matching gel was waived. Instead, two small pieces of adhesive tape were put on the borders of the fiber holder. The tape pieces served as 30 µm distance holders and prevented scratching when sliding the fiber holder towards the lens. As dispersive element, a volume phase holographic grating (VPHG) was used. The shape of the diffraction grating made of gelatin is 118 mm diameter circular while the entire VPHG component appears with a square shape of 122 mm side. The grating is optimized to cover a wavelength range from 350 to 900 nm. Finally, the diffracted transmitted light is focused on the image plane of a custom-made CCD (charge coupled device) camera. The detector is a large area back-illuminated chip (CCD213, e2v, Chelmsford, UK) with 4096 × 4112 pixels and 15 µm pixel size. In the range from 400 to 800 nm, the detector shows a quantum efficiency of approx. 90 %, at 900 nm it is approx. 60 %. The spectrograph is capable of covering a wavelength range from 350 to 900 nm with a linear dispersion of 0.13 nm / pixel. However, for sources with a spectral energy distribution larger than one octave, the overlap of second order signals must be suppressed by use of order separating filters. A description of the camera detector and its readout configuration can be found in [24]. In the standard configuration, approx. 30 s are needed to read out the complete CCD chip. The control software saves the raw signal as FITS file [25] and visualizes it by using the free available SAOImage DS9 software (Smithsonian Astrophysical Observatory, USA, available at http://ds9.si.edu). This viewer is useful to perform quick-look checks, e.g. to verify the presence of a suitable Raman signal by plotting a profile along the wavelength axis. For full data reduction and visualization, the raw data file is processed with the open source software P3D [26, 27] (available at https://p3d.sourceforge.io/). The main features of the software are: (I) The exact trace of the spectrum corresponding to each fiber is determined. Due to optical aberrations, the signal traces of the fibers are not parallel to pixels of the detector. The related deviations are determined by continuous signal traces on the detector generated when illuminating all fibers of the bundle with white light. (II) The dispersion across the detector has an arc-shaped characteristic. This effect is commonly known as spectral smile or keystone [28]. By use of a Ne-Lamp or another light source with discrete emission lines, a wavelength calibration is applied, i.e. the wavelength corresponding to each pixel is determined as a polynomial solution. (III) Every signal trace is assigned to the spatial position of the corresponding fiber in the matrix. Finally, a data cube with two spatial axes and one spectral axis is provided as a FITS file. (IV) A viewer module of the software allows to inspect the cube in different ways (display of spatial maps at a chosen wavelength, plot of spectra of selected regions) and to perform simple data analysis tasks such as flux measurements and spectral line fits.

## 3. Epoxy Skin Phantoms

To characterize the setup, phantoms based on epoxy resin (epoxy casting resin "waterclear", R&G Faserverbundwerkstoffe, Germany) were prepared. Since the setup is intended to measure skin samples, the absorption and scattering properties of the phantoms were aligned to human dermis [29] by adding proper amounts of $TiO_2$ and black epoxy paste as scatter and absorber, respectively. After curing, the surface was polished, finally using a sand paper with 15 μm grain size. The scattering and absorption coefficients of different phantoms were measured (Lambda 900 UV/VIS/NIR Spectrometer with integrating sphere; Perkin Elmer, USA) to $\mu_s' = 2.1 \pm 0.3$ mm$^{-1}$ and $\mu_a = 0.05 \pm 0.01$ mm$^{-1}$, respectively, at 785 nm wavelength.

## .4. Characterization of the Experimental Setup

### 4.1 Image Acquisition Head

Raman imaging with IFS ideally requires a homogeneous laser illumination of the entire image field during the recording process. To verify the conditions of our setup, the illumination of a sample image was simulated by software (OpticStudio, Zemax LLC Delaware, USA). The pseudo color image of the simulation (Fig. 3, left) shows that most of the excitation energy is contained in 50 μm diameter spots at the focus positions of every micro-lens. Because of the filling-factor of 65 %, spaces between the micro-lenses are also weakly illuminated. At least for non-scattering samples, the remaining 35 % do not contribute to excitation, especially since Raman signals arising from the gaps between the lenses are not coupled into the fibers at the other end of the image acquisition head. Due to optical aberrations, the illumination geometrical spot sizes are not exactly constant across the focal plane of the MLA, but increase slightly near the edge of the field of view. The simulation reveals that excitation energy within the spots drops from the MLA center to the penultimate rows by approximately 20 %. From the penultimate row to the last row there is an additional drop of 40 %.

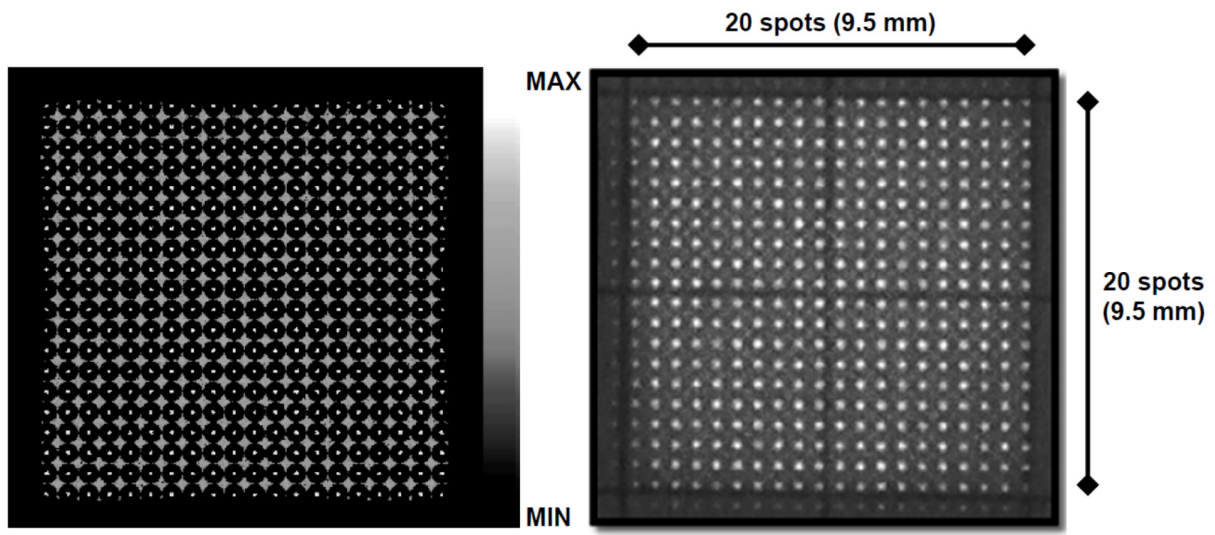

**Fig. 3 left** Simulated excitation intensity distribution at the sample side. The grey scale is logarithmic. **Right** Unprocessed (linear) camera image of a scale paper placed on the top of the image acquisition head. The spots show the excitation laser light.

For comparison, Figure 3, right shows a real camera image of the excitation spot received by switching on the laser and putting a piece of scale paper on the top of the image acquisition head. Due to scattering of the paper, the spots appear larger than predicted by the simulation. To a certain extent the intensity impression is an artefact of the camera optics. The actual intensities were measured using a laser power meter in combination with a shadow mask. The total power of the excitation laser at the exit of the square fiber was 400 mW. Except for the four border rows, intensities of approximately 1 mW / 0.25 mm$^2$ were measured. Obviously no significant loss of laser light occurs within the excitation pathway. However, at the border rows the intensities drop to approximately 0.6 mW / mm$^2$, which is in accordance with the simulation. Obviously, the shaft of the MLA holder causes vignetting. In conclusion, an image field of 18 × 18 pixels is approximately homogeneously illuminated to within ± 10 %. It should be noted that the edges of the MLA's do not run exactly between mircrolenses. Thus, the central microlenses are not precisely in the center of the holder and hence, the intensity gradients are not perfectly symmetric.

*4.2 Spatial Resolution for Skin Samples*

If a scattering sample like skin is examined, homogenous illumination of the surface causes a decrease in spatial resolution because multiple scattering in the sample leads to crosstalk between the pixels: Photons originating from illumination of a specific pixel also contribute to the Raman signal at neighboring pixels. To estimate this effect, a Monte Carlo simulation using scattering and absorption coefficients of human skin $\mu'_s$ = 1.46 mm$^{-1}$ and $\mu_a$ = 0.044 mm$^{-1}$ [29] has been performed, see Figure 4. Each fiber of the matrix captures light arising from certain sections of the sample surface. From the magnification factor of the image acquisition head, the aberrations, and the fiber core diameters, it was estimated that these sections are spots with approximately 100 μm diameters.

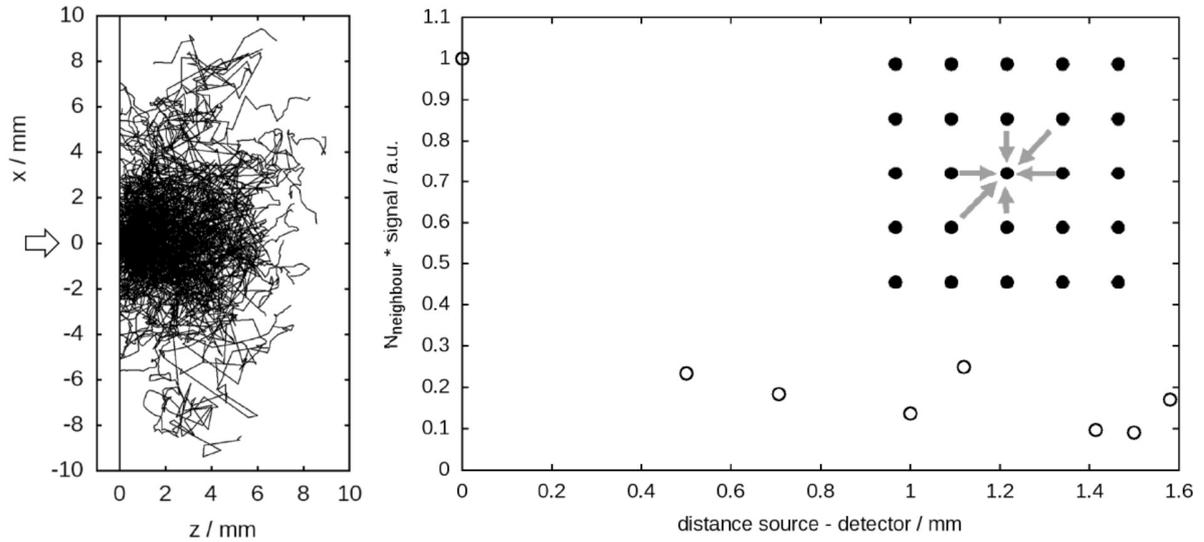

**Fig. 4** Monte Carlo simulation of multiple scattering in skin. **Left** XZ projection showing the traces of photons impinging the skin sample at zero position. **Right** Contributions of signals arising at neighbored excitation spots.

Figure 4 right shows the contributions of adjacent pixels in a 0.5 mm raster. A signal strength of 1.00 is assumed if only one pixel at detection position (pixel at center of the insert) would be illuminated. If also illuminating the neighboring pixels, additional Raman-scattered photons will reach the pixel in the center. Four neighboring pixels in 0.5 mm distance add to 0.23 signal strength. Four neighboring pixels in 0.71 mm distance add to 0.18 signal strength. At 1.1 mm distance there are 8 pixels contributing in total to 0.25 signal strength. Adding up all neighboring contributions, it turns out that approximately half of the signal detected on a certain position comes from scattered light originating from adjacent pixels.

The resulting effect on the lateral resolution is shown in Figure 5 for the same human-skin model, where a homogeneously distributed Raman active component $c(r)$ is restricted laterally to a half space ($c(r) = 0$ for $x < 0$). This example is intended to correspond to an experiment where the boundary of a large-area tumor region is examined. It is assumed that the cancerous region has a specific Raman signal that is not present in the healthy part. This assumption is a simplification.

In reality, cancer is not indicated by a clear presence or absence of a definite Raman peak, but rather by slightly changes of the spectrum. Nevertheless, the assumption is meaningful to depict the consequences of channel crosstalk. Figure 5 shows two situations: (I) All spots are illuminated simultaneously, as it is the case for our setup (black circles) and (II) only one spot located at the measurement position is illuminated, as it is the case for a pointwise scanning Raman microscope. For our setup, the transition of the Raman signal at the boundary ($x = 0$) is remarkably broadened to more than 1 mm, whereas a single spot setup would result in a resolution far better than 0.5 mm.

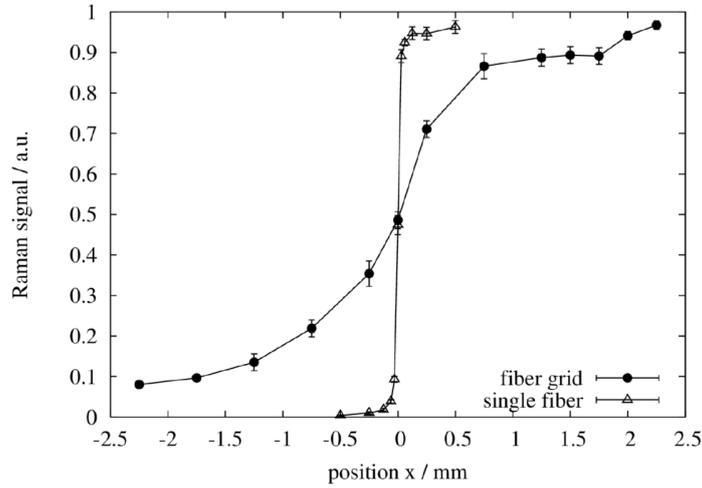

**Fig. 5** Calculated Raman signal (normalized to maximum value) for a large-area Raman source restricted to x > 0 for illuminating all pixels (fiber grid) and a stepwise illumination of only one pixel (single fiber). The abscissa is the position of the receiving fiber relative to the boundary of the Raman source.

The Monte Carlo simulation also provides information about the signal distribution from different depths. For measurements on skin tissue it is important to know the spatial distribution of Raman-scattering processes detected by the instrument. The Raman signal can be described with an integral over the probe volume which contains a weighting function $w(r)$, as shown in equation (1).

$$S_{\text{Raman}} = \int_r w(r)\sigma_{\text{Raman}} c(r) d^3 r \tag{1}$$

$c(r)$ is the spatially-dependent concentration of the Raman active molecule, $r$ the spatial position, and $\sigma_{\text{Raman}}$ the Raman cross section [30]. The function $w(r)$ is a measure of the sensitivity of the setup for Raman scattering at location $r$.

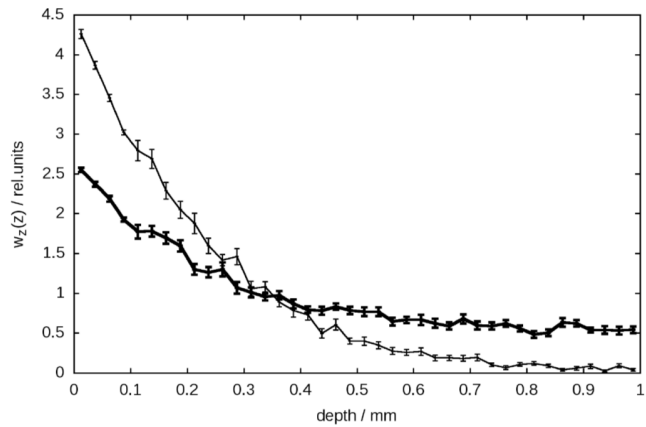

**Fig. 6** Monte Carlo simulation of the Raman signal contribution from different depths of skin. $w_z(z)$ is the laterally integrated weight function (see Eq. 1). Bold: Illumination of all excitation spots. Thin line: Illumination at the detection spot only. The functions are normalized to unit area.

Figure 6 shows the depth dependence of the weight function $w(r)$ for Raman signal detected at one spot. The thin curve shows the signal contributions from Raman scattering at different depths as they would be received if only the detection pixel (spot) would be illuminated. In this case, most of the signal would arise in a range from the surface to approximately 0.3 mm depth. The bold curve shows the corresponding contributions if all pixel (spots) are illuminated. In comparison to single pixel illumination, the simultaneous illumination of all pixels results in the detection of Raman-scattering processes in larger depths due to crosstalk as discussed for Figure 4. In the case of real skin, the most intense Raman signals arise from the epidermis and the top of the papillary dermis (approx. 50 μm), but can be enhanced down to approx. 200 μm using the optical clearing technique [31].

*4.3 Homogeneity of the Raman Images*

The epoxy phantoms were used to examine the spatial signal intensity distribution of the setup. The homogeneity of the samples, more precisely of the signals arising from the phantoms, was verified by using a scanning single channel Raman spectrometer (Laser- und Medizin-Technologie GmbH, Germany; excitation wavelength 785 nm) [5]. At various positions of a phantom the spectra were measured within a square of 2 cm border length. The variation of all unprocessed spectra was less than ± 5 %, proving the homogeneity of the phantoms (Fig. 7).

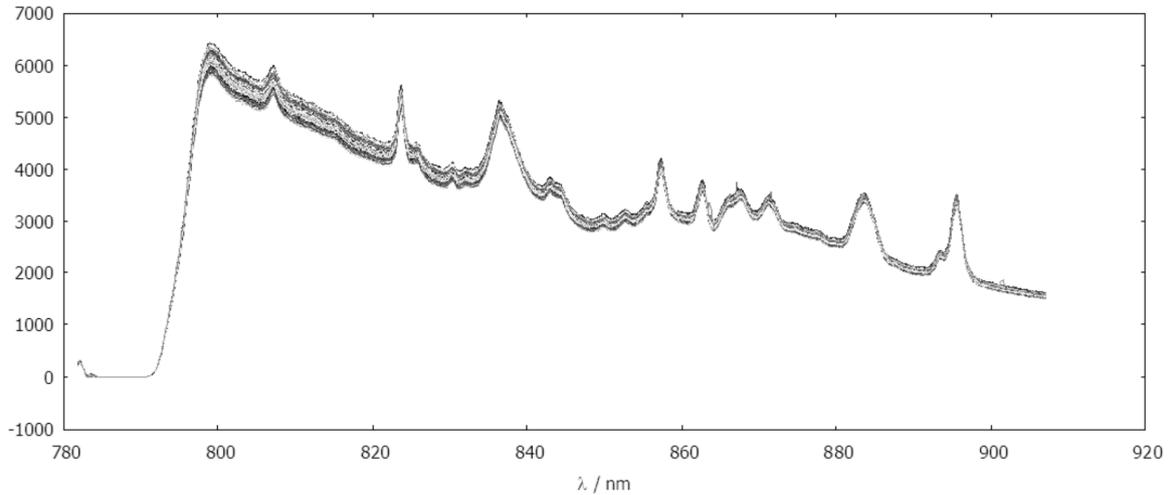

**Fig. 7** Raw spectra of an epoxy phantom measured with a single channel spectrometer at different positons within a square of 2 cm border length.

Although homogeneous phantoms are used as samples, the received signal distribution was not uniform. Figure 8 shows the average intensity distributions of the raw spectra of a phantom, which appears as a dome-like shape with center intensities that are approximately four times higher than at the corners.

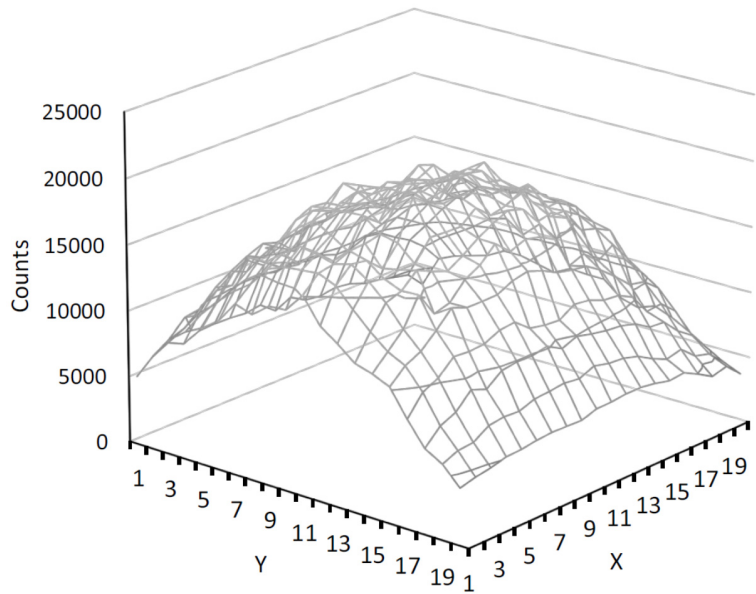

**Fig. 8** Intensity distribution when measuring a homogeneous phantom. Instead of a flat distribution, the result shows a dome shape.

The intensity differences can be associated with the spectrograph's properties (I), the characteristics of the image acquisition measurement head (II), and the scattering properties of the sample (III).

The spectrograph's sensitivity (I) depends on the positions of the fibers located in the V-grooves. Due to the design of the spectrograph, light emerging from fibers at the border positions of the V-groove holder is vignetted to a certain amount. The flat field correction uncovers this characteristic. To perform the flat field measurement, the fiber bundle matrix (Fig. 2) is removed from the image acquisition head and homogeneously illuminated by using an integrating sphere. The fibers in the matrix are sorted with regards to their positions in the V-groove. Fibers at the positions (X,Y) = (1,1) and (20,20) are located at the borders of the V-groove holder, whereas the fibers at (10,20) and (11,1) are in the center. As expected, the spatial sensitivity results in a half pipe shape showing intensity. Minima at the corners are probably caused by some vignetting due to the limited size of the available integrating sphere (data not shown). A further potential source of inhomogeneities is the distribution of the excitation intensity (II). As shown above, the excitation intensity declines from the center to the border rows. A further reason for the dome-shape is the scattering of the samples (III). As described above, scattering leads to significant signal contributions from neighboring spots. Indeed, fibers in the center of the matrix receive the largest contribution from neighbors. When moving from the center to the borders, the numbers of neighbors and, accordingly, the signal decreases. Assuming homogeneous illumination and absorption and scattering coefficients of skin, a calculation shows that the relative Raman intensity drops from 1 to 0.6 when moving 10 fibers diagonally from the corner to a center. Finally, a combination of (I), (II) and (III) leads to the dome-like shape shown in Figure 7. Since the actual absorption and scattering coefficients, hence the intensity distribution, differ from sample to sample, a universal correction function cannot be applied. Therefore, it was decided to include normalization into the preprocessing of spectra. Fluorescence background was removed using a 6th order polynomial fitted to each spectrum [32]. Subsequently, the standard normal variate (SNV) normalization was applied [33]. It should be mentioned that the used laser unit includes two excitation lasers with 1 nm gap. Thus, the setup is capable to remove background by use of shifted excitation differential Raman spectroscopy (SERDS) [18, 34]. However, the SERDS option was not used in this work, since Raman spectra reconstructions based on SERDS curves considerably differ from common Raman spectra, making a comparison to previously published results difficult. Further drawbacks of SERDS are the negative impact of photobleaching and the extension of the measurement time. It was therefore decided to apply a "classic" polynomial-based background removal algorithm. As shown in Figure 9, highly uniform spectra result from the preprocessing. From neighboring pixels it was estimated that the variance between the spectra is approximately only twice the variance due to noise.

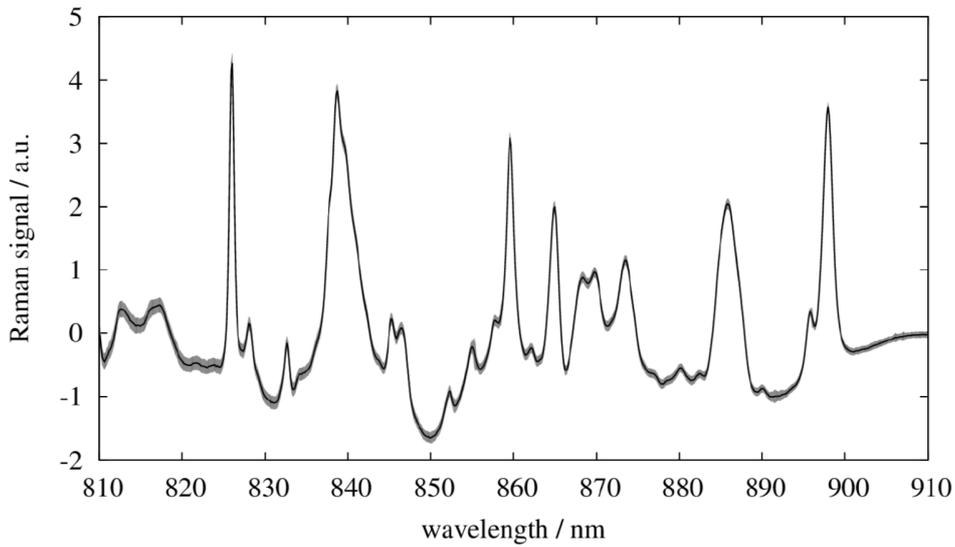

**Fig. 9** Average and range of variation (± σ) for normalized Raman spectra of a homogeneous epoxy sample.

To visualize the variations between the spectra of the Raman image, a principal components analysis (PCA) of the set of 400 spectra of an image was performed. Each spectrum of the image is represented as a linear combination of the orthonormal set of principal components $p_m(\lambda)$:

$$y_i(\lambda) = \overline{y}(\lambda) + \sum_m s_i^{(m)} p_m(\lambda) \tag{2}$$

where $\bar{y}$ is the average spectrum of the image. Most of the variance is described by a small subset of the principal components. Figure 10 shows the first three principal components ($m$ = 1, 2, 3) for an epoxy phantom. There are variations between the spectra in the wavelength region of 810 to 840 nm, which are probably due to slight variations of filter-transmission spectra. Near the filter edge, the transmission curve shows fringes. The angle of incidence and therewith the effective filter response varies slightly with the position of each pixel.

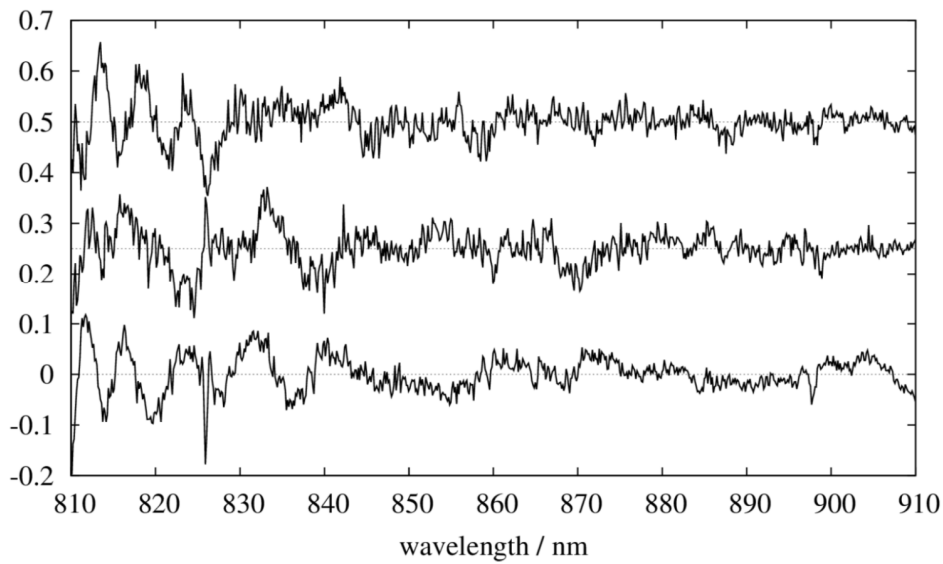

**Fig. 10** Principal component analysis: First principal components $p_1$ (bottom), $p_2$ (center), and $p_3$ (top) describing the largest variance between spectra (the offsets are introduced for clarity).

Except for two small peaks at 825 nm and 900 nm, the principal components do not contain Raman bands of the epoxy sample, therefore it seems reasonable that they only describe the instruments variation of spectral sensitivity.

*4.4 Comparison of Multichannel Setup with Single Channel Spectrograph*

The measurements of epoxy phantoms and the PCA analysis described above were also used to evaluate the influence of the Raman-image inhomogeneities in a medical application. A single channel spectrograph (constructed at the Laser- und Medizin-Technologie GmbH, Germany) has been successfully applied to examine biopsy tissue samples [5]. For this setup a Raman image was realized using a motorized XY stage. The spectral resolution was 0.25 nm, i.e. similar as for our instrument. Partial least squares discriminant analysis (PLS-DA) has been used to discriminate normal from precancerous tissue. PLS-DA results in a discriminant function which assigns each tissue spectrum $g_k$ a scalar value $d_k$ by means of a scalar product that contains a weight function $b$ and an average spectrum $\overline{g}$ (Fig. 11), the tissue type is inferred from the sign of $d_k$:

$$d_k = \langle g_k - \overline{g}, b \rangle \quad \begin{cases} d_k < 0 \Rightarrow k \in class\ 1\ \text{(normal tissue)} \\ d_k \geq 0 \Rightarrow k \in class\ 2\ \text{(precancerous tissue)} \end{cases} \quad (3)$$

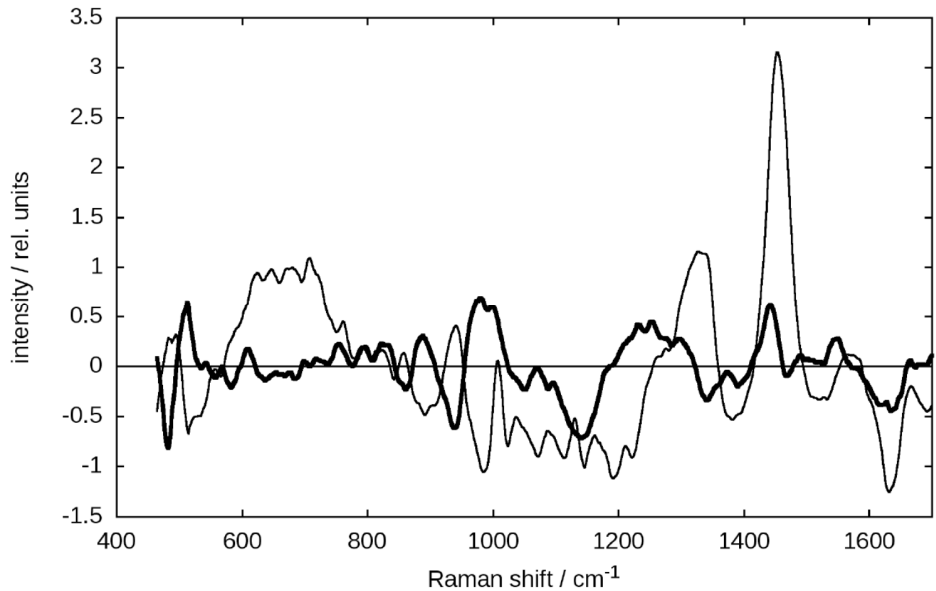

**Fig. 11** Average tissue spectrum $\overline{g}$ (thin line) and weight function $b$ (thick line) of the PLS-DA.

The weight function $b(\lambda)$ determines which parts of the spectrum are used for discrimination of tissue types. As $b(\lambda)$ results in positive and negative values, equation 3 also evaluates differences of spectral values. To estimate the influence of the Raman-image inhomogeneity, the tissue spectra $g_k$ are modified by the first principal component of the PCA analysis of the epoxy phantom, which describes the largest variation of the Raman-image inhomogeneity (inclusion of more principal components does not alter the result substantially):

$$g_k(\lambda) \rightarrow g_k(\lambda) + s_k^{(1)} p_1(\lambda) \Rightarrow d_k \rightarrow d_k + s_k^{(1)} \langle p_1, b \rangle \qquad (4)$$

For some spectra, the modification leads to a change of the sign of $d_k$ so that the quality of the discriminant analysis, characterized by its sensitivity and specificity [35] is reduced.

In principle, each tissue spectrum $g_k$ can occur on any position of the Raman-image. Therefore the coefficient $s^{(1)}$ can attain any value found in the PCA analysis of the Raman-image inhomogeneity. It follows that for each $d_k$ a distribution of values with a standard deviation given by

$$\sigma(d) = \sigma(s^{(1)}) \langle p_1, b \rangle \qquad (5)$$

is found. Assuming a normal distribution for each $d_k$, average values for sensitivity and specificity have been derived according to equation 4, as shown in Figure 12 as a function of $\sigma(d_k)$. The standard deviation $\sigma(s^{(1)})$ follows from the PCA analysis of the epoxy phantom. With equation 5, a value of $\sigma(d_k)$ = 0.003 follows. One can conclude from Figure 12 that the Raman-image inhomogeneity will have no significant effect on the tissue discrimination in this application.

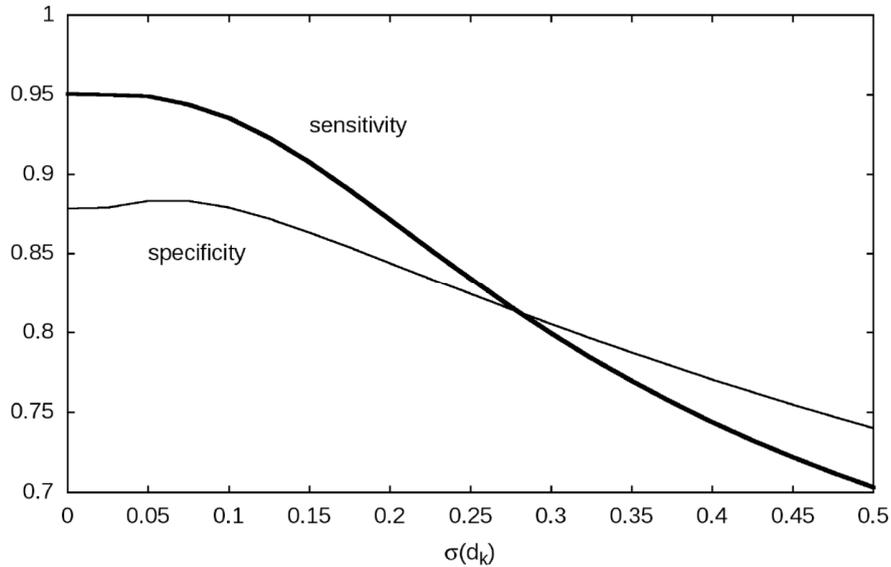

**Fig. 12** Reduction of sensitivity and specificity for discrimination of two tissue classes (normal, precancerous).

## 5. Results and Discussions

*5.1 Nevus at forearm in vivo*

The setup was already tested on porcine skin samples *ex vivo* [18]. To verify the capability for human skin samples *in vivo*, a forearm of a volunteer was placed on the top of the measurement head. Figure 13, left delineates the test area as a skin imprint of the measurement head. A nevus with approximately 3 mm diameter can be seen in the upper left. The recording time was 2 min. Due to the fluorescence of the melanin the position of the nevus can easily be seen on the basis of the signal strength of the raw data (not shown). However, the aim of the experiment was to identify the nevus only with the aid of Raman spectra without any fluorescence background. Thus, all fluorescence background was removed by a polynomial fit followed by a normalization of the spectra. To the resulting background-free Raman spectra PCA was applied. Figure 13, center shows the Raman spectra inside and beyond the position of the nevus *in vivo*. Figure 13, right shows the difference of the principal components 2 and 3 (PC2 - PC3) received from the PCA.

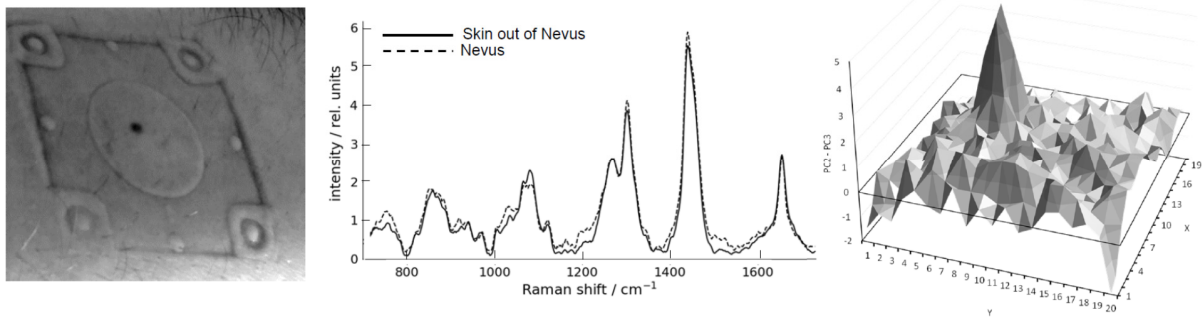

**Fig. 13, left** Nevus on a forearm. The circular skin imprint caused by the image acquisition head has a diameter of 20 mm. **Center** Raman spectra at the position of the nevus (dotted curve) and beside it (continuous curve). **Right** The difference of the principal components PC2 and PC 3 of the Raman spectra reflects the position of the nevus.

The nevus can be clearly seen as peak. Its diameter of the peak's footprint is approximately 7 pixels diameter which corresponds to 3.5 mm, i.e. the Raman image reflects the actual size of the nevus at the camera image. Thus, the achieved resolution of 0.5 mm matches the distance of the fibers in the matrix.

*5.2 Discrimination of skin regions in vivo*

As a test case Raman-spectral maps of six skin regions (forearm, volar forearm, palm, thumb, leg, foot) have been measured on four volunteer persons. The aim of this experiment was to verify whether the system is able to distinguish between different body parts. Following preprocessing as described above, principal component analysis (PCA) was applied to the total set of spectra [36]. The scores of the first two principal components are shown in Figure 14.

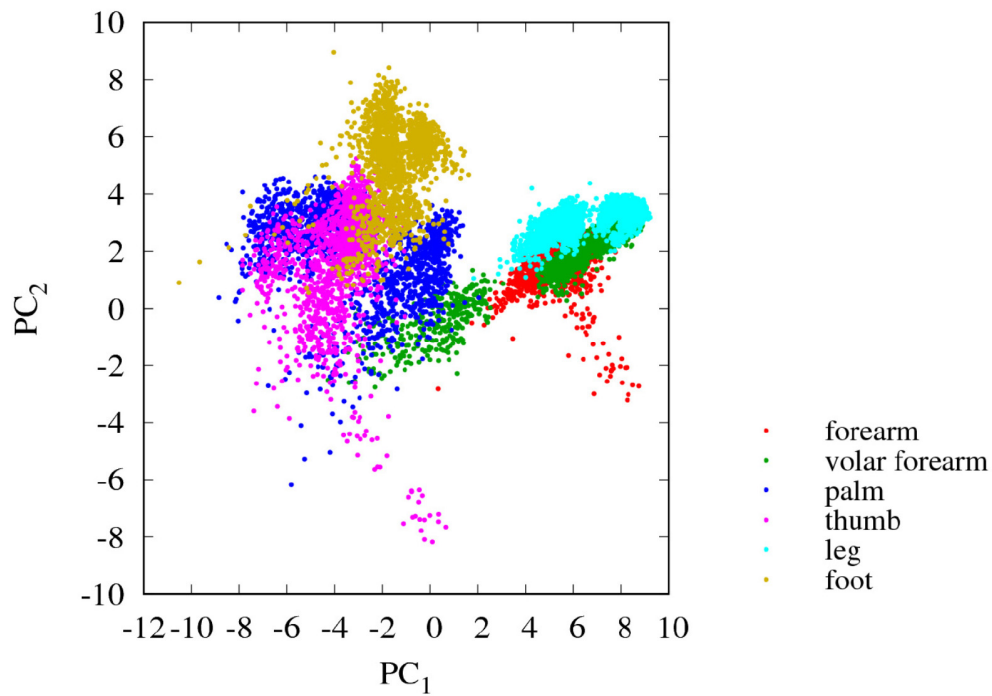

**Fig. 14** Scores of two principal components of skin Raman spectra from four persons in vivo. Colors according to analyzed skin of different body sites.

As can be seen from Figure 14 the skin regions differ in their spectral features. For example, leg (cyan points) and foot (light brown points) are well separated and can be easily discriminated. To investigate, whether one can discriminate the skin regions using more spectral information, an unsupervised clustering algorithm was applied [37]. Each spectrum was reduced to the scores of four principal components. Inclusion of more than four components did not improve the result. The distribution of spectra on six clusters show that the spectra of forearm, palm and thumb and spectra of forearm and leg are not clearly separated by the cluster algorithm (Table 1). Only the spectra of foot skin are clearly distinct, probably due to the very large thickness of the stratum corneum in this case. The main reason for the missing discriminability of the skin of different body sites is the inter-individual variance of the skin. The thickness of skin layers, as well as microcirculation of blood in the papillary dermis and the melanin concentration varies for different body sites. The concentration of carotenoids, which are most concentrated in the stratum corneum, is an individual parameter, depending on the skin area [38] and volunteers' lifestyle [39]. This is apparent when the cluster algorithm is applied to all spectra of a single person *in vivo*, as shown in Table 2 for an example. Here most of the skin spectra of a body site are assigned to a single cluster.

|           | Dorsal forearm | Volar forearm | Palm | Thumb | Leg | Foot |
|-----------|----------------|---------------|------|-------|-----|------|
| Cluster 1 | 1192           | 703           | 0    | 0     | 798 | 0    |
| Cluster 2 | 4              | 335           | 701  | 5     | 1   | 0    |
| Cluster 3 | 0              | 0             | 591  | 1125  | 0   | 26   |
| Cluster 4 | 0              | 0             | 50   | 349   | 0   | 3    |
| Cluster 5 | 1              | 497           | 0    | 0     | 800 | 0    |
| Cluster 6 | 0              | 0             | 2    | 4     | 0   | 1544 |

**Table 1** Partition of skin spectra *in vivo* from four persons into six clusters obtained by hierarchical clustering. Unsupervised hierarchical clustering using a minimum variance method [37].

|           | Dorsal forearm | Volar forearm | Palm | Thumb | Leg | Foot |
|-----------|----------------|---------------|------|-------|-----|------|
| Cluster 1 | 374            | 8             | 0    | 0     | 0   | 0    |
| Cluster 2 | 0              | 322           | 0    | 0     | 91  | 0    |
| Cluster 3 | 0              | 0             | 298  | 8     | 0   | 27   |
| Cluster 4 | 18             | 0             | 0    | 329   | 0   | 0    |
| Cluster 5 | 6              | 69            | 0    | 44    | 308 | 0    |
| Cluster 6 | 0              | 0             | 1    | 0     | 0   | 373  |

**Table 2** Partition of skin spectra *in vivo* into six clusters obtained by hierarchical clustering. Spectra from one person.

For discrimination of skin of different body sites without reference to individual properties a more refined discrimination analysis including, for example, supervised learning and feature extraction is necessary [40]. Eventually, inter-area differences (Fig. 14) exist and should be considered by comparison between cancerous and healthy skin samples. However, this is beyond the scope of this pilot study, and would require the investigation of many more samples: clearly the objective of a detailed follow-up study.

*5.3 Validation of multiplex Raman for cancer diagnostics*

Raman spectra of human skin cancerous and healthy tissue were recorded *ex vivo* and investigated as a possible future diagnostics for skin cancer *in vivo*. Pairs of healthy and cancerous skin biopsy samples were supplied from the Department of Dermatology, Charité. The samples were taken from various patients after surgery and are listed in Table 3. The experiments were approved by the Ethics committee of the Charité-Universitätsmedizin, Berlin (EA1/340/16) and conducted according to the declaration of Helsinki. All volunteers gave their written informed consent. Most of the samples were cylindrical since taken by core biopsy. After removal from the patient, samples 4a and 4b were stored at 6 °C and measured one day later. All other samples were at first frozen at -30 °C and stored. For the transportation from the hospital to the location of the Raman spectrograph, an ice-filled styrofoam box was used. Before starting the Raman measurements, the samples were slowly de-frozen in a refrigerator at 6 °C.

| Sample # | Origin | Suspicion | Finding | Size | Sex of Patient | Age | Frozen |
|---|---|---|---|---|---|---|---|
| 1a | nasal wing | normal | - | ⌀ 10 mm | f | 75 | yes |
| 1b | " | BCC | not confirmed | ⌀ 4 mm | " | " | yes |
| 2a | forehead | normal | - | ⌀ 10 mm | m | 75 | yes |
| 2b | " | SCC | AK grade 2-3 | ⌀ 4 mm | " | " | yes |
| 3a | cheek | normal | - | ⌀ 4 mm | m | 74 | yes |
| 3b | " | BCC | SCC | ⌀ 10 mm | " | " | yes |
| 4a | head | normal | - | ⌀ 4 mm | m | 76 | no |
| 4b | " | SCC | SCC | ⌀ 4 mm | " | " | no |
| 5a | upper arm / shoulder | normal | - | ⌀ 4 mm | m | 79 | yes |
| 5b | " | BCC | BCC | ⌀ 4 mm | " | " | yes |
| 6a | shoulder | normal | - | ⌀ 10 mm | m | 73 | yes |
| 6b | external ear | BCC | BCC | ⌀ 4 mm | " | " | yes |
| 7a | lower leg | normal | - | ⌀ 10 mm | m | 77 | yes |
| 7b | " | SCC | Basosquamous carcinoma | ⌀ 6 mm | " | " | yes |
| 8a | temple | normal | - | ⌀ 2 mm | w | 91 | yes |
| 8b | " | BCC | BCC | ⌀ 2 mm | " | " | yes |

**Table 3** Biopsy skin samples. BCC: basal cell carcinoma, SCC: squamous cell carcinoma, AK: actinic keratosis. Two consecutive samples originate from one volunteer, where the first sample is normal and the second sample is cancerous.

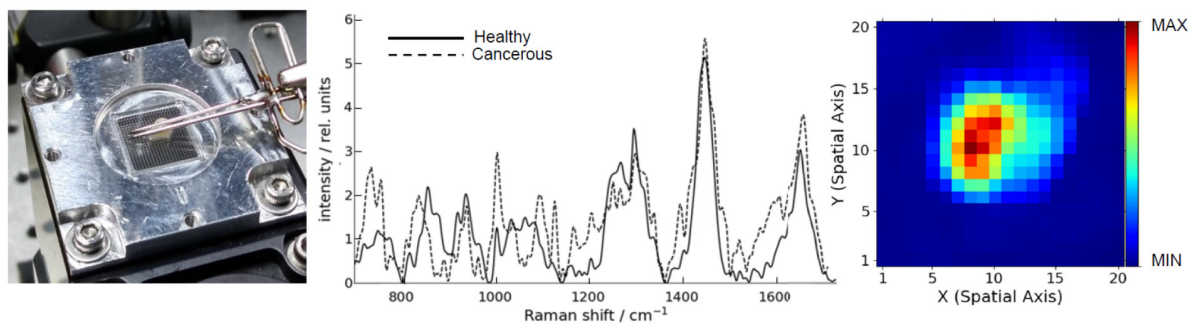

**Fig. 15, left** Human skin biopsy sample from external ear at the top of the image acquisition head (sample 6b listed in Table 3). **Center** *Ex vivo* Raman spectra of cancerous human skin tissue at position (8,10) (dotted line) in comparison with the corresponding healthy biopsy sample (continuous line). **Right** Intensity of the Raman signal at 1448 cm$^{-1}$ in color levels.

Figure 15 left shows the biopsy sample 6b (Table 3) placed at the top of the image acquisition head. Two parallel needles were used to hold the sample by clamping or skewering. Before and during the measurement, the sample was covered by a lid containing a wet paper tissue to avoid drying. For some of the samples, it was necessary to increase the excitation intensity. The maximum output power of the used laser source is 500 mW. Taking into account the MLA fill factor and losses in the excitation pathway limits the excitation intensity to approx. 0.8 mW / pixel when illuminating all 400 lenses, which is safe for *in vivo* application. For the samples replacing the 600 μm square core fiber by a 300 μm square core fiber, the excitation power is concentrated to 100 lenses, i.e. the excitation power increases four-fold.

Since the number of available biopsy samples was too low to perform any reliable refined discrimination analysis, it was decided to assess the results on the basis of the average spectra. Biopsy samples of normal skin from four patients and samples of basal-cell carcinoma (BCC) from three patients have been measured, summarized and normalized. The available samples were either entirely affected with cancer or entirely healthy. Thus, within a sample, the spectral patterns were quite similar at all positions. Figure 16 shows the resulting ranges of Raman-spectral values for both types of skin. The comparison of the Raman spectra reveals the following differences between normal skin and BCC:

- A narrower line shape for BCC at the Amide I – region (1640 – 1680 cm$^{-1}$)
- The intensity ratio of region (1220 – 1290 cm$^{-1}$) (Amide III – region) to region (1290 – 1360 cm$^{-1}$) decreases for BCC.
- Decreasing bands in the region (830 – 980 cm$^{-1}$) for BCC.

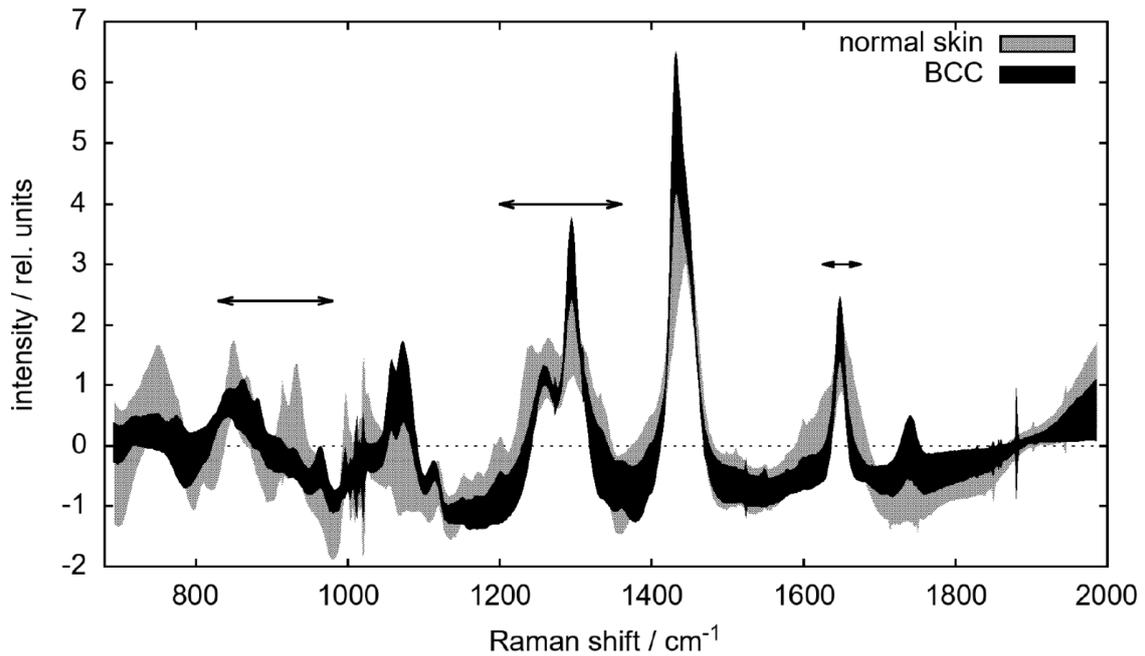

**Fig 16** Range of spectral values (average ± $\sigma$) for normal skin (gray) and BCC biopsies (black). Wavenumber region marked with arrows are discussed in the text.

These differences are in accordance with published results received from a single channel Raman setup [41]. However, with our setup, far more pixels could be measured in shorter time. In [41], 10 min measurement time was applied to record the Raman spectrum of one single 100 μm diameter spot. In this work, the Raman spectra of 100 spots were recorded within 2 min. Even though the experimental conditions differ to a certain extend (up-to-dateness of the hardware, excitation wavelength, excitation power), the achieved measurement speed increase from 10 min / spot to 1.2 s /spot is striking and documents the capability of IFS even in the field of optical cancer diagnosis.

## 6. Conclusion and Outlook

By transferring IFS from astronomy to imaging Raman spectroscopy a setup was realized that is capable to measure the Raman spectra of 400 pixels of a 1 cm$^2$ sample simultaneously without any scanning procedure. Within this work the applicability of this setup for examination of human skin patches was investigated. Monte Carlo simulation yielded that a spatial resolution of approximately 1 mm can be achieved when illuminating the whole sample area with the excitation light. This result was experimentally confirmed by recording the Raman image of a nevus *in vivo*. Finally, biopsy samples of cancerous (BCC, SCC and AK) and healthy parts of skin were examined. The changes in Raman spectra match literature values received with a classic single channel spectrometer. Although the few small biopsy samples available for this experiment did not allow to directly localize distinct malign and benign tissue on a single sample, as we would require to actually prove the capability of this method to determine resection margins as the ultimate goal for surgery, we were able to demonstrate the plausibility by means of in vivo measurements of a nevus. Clearly, the next step must be a detailed clinical study to validate this finding on the basis of statistically meaningful samples and larger pieces of tissue. However, already the results of our pilot study show great promise for the development a minimal-invasive optical medical device for detecting cancerous tissue *in vivo* in the future. Currently, a follow-up project is being undertaken to examine the implementation of IFS in a medical endoscope.

## 7. Conflicts of Interest

The authors declare that there are no conflicts of interest.

## 8. Acknowledgement

The authors acknowledge support from the German Federal Ministry of Education and Research (BMBF) VIP program, "Multiplex-Raman-Spektroskopie" (MRS), grant no. 03V0843.

## 9. References

[1] C. Krafft, J. Popp, "The many facts of Raman spectroscopy for biomedical analysis," *Anal. Bioanal. Chem.* **407**, 699-717 (2015) [doi:10.1007/s00216-014-8311-9].

[2] C. Krafft, M. Schmitt, I. W. Schie, D. Cialla-May, C. Matthäus, T. Bocklitz, and J. Popp, "Label-Free Molecular Imaging of Biological Cells and Tissues by Linear and Nonlinear Raman Spectroscopic Approaches," *Angew. Chem. Int. Ed.* **56** 4392-4430 (2017) [doi:10.1002/anie.201607604].

[3] C.-S. Choe, J. Lademann, M. E. Darvin, "Depth profiles of hydrogen bound water molecule types and their relation to lipid and protein interaction in the human stratum corneum in vivo," *Analyst* **141**(22) 6329-6337 (2016) [doi: 10.1039/c6an01717g].

[4] H. Lui, J. Zhao, D. McLean, and H. Zeng, "Real-time Raman Spectroscopy for In Vivo Skin Cancer Diagnosis," *Cancer Res.* **72**(10), 2491-2500 (2012) [doi:10.1002/jbio.201400026].

[5] C. Reble, I. Gersonde, C. Dressler, J. Helfmann, W. Kühn, and H. J. Eichler, "Evaluation of Raman spectroscopic macro raster scans of native cervical cone biopsies using histopathological mapping," *J. Biomed. Opt.* **19**(2), 027007 (2014) [doi:10.1117/1.JBO.19.2.027007].

[6] X. Feng, A. J Moy, H. T. M. Nguyen, J. Zhang, M. C. Fox, K. R. Sebastian, J. S. Reichenberg, M. K. Markey, and J. W. Tunnell, "Raman active components of skin cancer," *Biomed. Opt. Express* **8**(6), 2835-2850 (2017) [doi:10.1364/BOE.8.002835].

[7] I. P. Santos et al., "Raman spectroscopy for cancer detection and cancer surgery guidance: translation to the clinics," *Analyst* **142**(17) 3025-3047 (2016) [doi: 10.1039/c7an00957g].

[8] I. A. Bratchenko, D. N. Artemyev, O. O. Myakinin, Y. A. Khristoforova, A. A. Moryatov, S. V. Kozlov, V. P. Zakharov, "Combined Raman and autofluorescence ex vivo diagnostics of skin cancer in near-infrared and visible regions," *J. Biomed. Opt.* **22**(2) 27005 (2017) [doi: 10.1117/1.JBO.22.2.027005].

[9] R. W. Smits, S. Koljenović, J. A. Hardillo, I. ten Hove, C. A. Meeuwis, A. Sewnaik, E. A. Dronkers, T. C. Bakker Schut, T. P. Langeveld, J. Molenaar, V. N. Hegt, G. J. Puppels, R. J. Baatenburg de Jong, and D. W. Eisele, "Resection margins in oral cancer surgery: Room for improvement," *Head Neck*, **38**, E2197-E2203 (2016) [doi:10.1002/hed.24075].

[10] J. J. Christophel, A. K. Johnson, T. L. McMurry, S. S. Park, and P. A. Levine, P. A. (2013), "Predicting positive margins in resection of cutaneous melanoma of the head and neck," *The Laryngoscope*, **123**, 683-688 (2013) [doi:10.1002/lary.23799].

[11] S. J. Erickson-Bhatt, R. M. Nolan, N. D. Shemonski, S. G. Adie, J. Putney, D. Darga, D. T. McCormick, A. J. Cittadine, A. M. Zysk., M. Marjanovic, E. J. Chaney, G. L. Monroy, F. A. South, K. A. Cradock, Z. G. Liu, M. Sundaram, P. S. Ray, and S. A. Boppart, "Real-time Imaging of the Resection Bed Using a Handheld Probe to Reduce Incidence of Microscopic Positive Margins in Cancer Surgery," *Cancer Res.*, **75**(18), 3706-3712 (2015) [doi: 10.1158/0008-5472.CAN-15-0464].

[12] P. D. A. Pudney, "In vivo Raman spectroscopy of skin," *Spectroscopy Europe* **27**(2), 14-16 (2015).

[13] J. Schleusener, V. Carrer, A. Patzelt, J. Lademann, and M. E. Darvin, "Surface determination of 3D confocal Raman microscopy imaging of the skin," *Laser Phys Lett.* **14**(12) 125601 (2017) [doi:10.1088/1612-202X/aa8bcb].

[14] Q. Li, X. He, Y. Wang, H. Liu, D. Xu, F. Guo, "Review of spectral imaging technology in biomedical engineering: achievements and challenges", *J. Biomed. Optics* **18**(10), 100901 (2013) [doi:10.1117/1.JBO.18.10.100901]

[15] C. Vanderriest, "Fiber-Optics Dissector for Spectroscopy of Nebulosities around Quasars and similar Objects," *PASP*, **92**, 858–862 (1980) [doi:10.1086/130764].

[16] M. Roth, "Introductory review and technical approaches," in: *3D Spectroscopy in Astronomy*, edited by: E. Mediavilla, E. Arribas, M. Roth, J. Cepa-Norgué, and F. Sánchez, Eds., pp. 1-39, Cambridge University Press, New York, USA (2010).

[17] R. Bacon et al., "The second-generation VLT instrument MUSE: science drivers and instrument design," in Ground-based Instrumentation for Astronomy, A. F. M. Moorwood, M. Iye, Eds., *Proc. SPIE* **5492**, 1145-1149 (2004) [doi:10.1117/12.549009].

[18] E. Schmälzlin, B. Moralejo, D. Bodenmüller, M. E. Darvin, G. Thiede, and M. M. Roth, "Ultrafast imaging Raman spectroscopy of large-area samples without stepwise scanning," *J. Sens. Sens. Syst.* **5**, 261–271 (2016) [doi:10.5194/jsss-5-261-2016].

[19] Y. Kumamoto, Y. Harada, H. Tanaka, and T. Takamatsu, "Rapid and accurate peripheral nerve imaging by multipoint Raman spectroscopy," *Sci. Rep.* **7**, 845 (2017) [doi:10.1038/s41598-017-00995-y].

[20] M. Okuno and H. Hamaguchi, "Multifocus confocal Raman microspectroscopy for fast multimode vibrational imaging of living cells," *Opt. Lett.* **35**(24), 4096-4098 (2010) [doi:10.1364/OL.35.004096].

[21] Y. Q. Li, S.-Y. Sasaki, T. Inoue, A. Harata, and Ogawa, "Spatial Imaging Identification in a Fiber-Bundle Confocal Fluorescence Microspectrometer," *Appl. Spectrosc.* **52**(8), 1111-1114 (1998) [doi:10.1366/0003702981944832].

[22] M. Brückner, K. Becker, J. Popp, and T. Frosch, "Fiber array based hyperspectral Raman imaging for chemical selective analysis of malaria-infected red blood cells," *Anal. Chim. Acta*, **894**, 76-84 (2015) [doi:10.1016/j.aca.2015.08.025].

[23] B. Moralejo, M. M. Roth, P. Godefroy, T. Fechner, S. M. Bauer, E. Schmälzlin, A. Kelz, R. Haynes, "The Potsdam MRS spectrograph: heritage of MUSE and the impact of cross-innovation in the process of technology transfer," in Advances in Optical and Mechanical Technologies for Telescopes and Instrumentation II, R. Navarro, J. H. Burge, Eds. *Proc. SPIE* **9912**, 991222 (2016) [doi:10.1117/12.2232539].

[24] B. Moralejo, E. Schmälzlin, D. Bodenmüller, T. Fechner, M. M. Roth, "Improving the Frame Rates of Raman Image Sequences Recorded with Integral Field Spectroscopy using Windowing and Binning Method", *J. Raman Spectrosc.*, **49**(2), 372–375 (2018) [doi:10.1002/jrs.5268].

[25] D. C. Wells, E. W. Greisen, R. H. Harten, "Fits : a flexible image transportation system," *A&A. Supp.* 44, 363-370 (1981).

[26] C. Sandin, T. Becker, M. M. Roth, J. Gerssen, A. Monreal-Ibero, P. Böhm, P. Weilbacher, "P3D: a general data-reduction tool for fiber-fed integral-field spectrographs," *A&A* **515** A35 (2010) [doi:10.1051/0004-6361/201014022].

[27] C. Sandin, P. Weilbacher, F. Tabataba-Vakili, S. Kamann, O. Streicher, "Automated and generalized integral-field spectroscopy data reduction using p3d," in Software and Cyberinfrastructure for Astronomy II, N. M. Radziwill, G. Chiozzi, Eds., *Proc. SPIE* **8451**, 84510F (2012) [doi:10.1117/12.926092].

[28] N. Yokoya, N. Miyamura, A. Iwasaki, "Preprocessing of hyperspectral imagery with consideration of smile and keystone properties", in Multispectral, Hyperspectral, and Ultraspectral Remote Sensing Technology, Techniques, and Applications III, A. M. Larar, H.-S. Chung, M. Suzuki, Eds., *Proc. SPIE* **7857**, 78570B (2010) [doi:10.1117/12.870437].

[29] S. L. Jacques, "Optical properties of biological tissues: a review," *Phys. Med. Biol.* **58**(11) R37-R61 (2013) [doi:10.1088/0031-9155/58/14/500].

[30] C. Reble, I. Gersonde, C. A. Lieber, J Helfmann, "Influence of tissue absorption and scattering on the depth dependent sensitivity of Raman fiber probes investigated by Monte Carlo simulations," *Biomed. Optics express*, **2**(3), 520-533 (2011) [doi:10.1117/1.3526367].

[31] A. Y. Sdobnov, V. V. Tuchin, J. Lademann, and M. E. Darvin, "Confocal Raman microscopy supported by optical clearing treatment of the skin-influence on collagen hydration," *J. Phys. D* **50**, 285401 (2017) [doi: 10.1088/1361-6463/aa77c9].

[32] A. Cao, A. K. Pandya, G. K. Serhatkulu, R. E. Weber, H. Dai, J. S. Thakur, V. M. Naik, R. Naik, G. W. Auner, R. Rabah, and D. C. Freeman, "A robust method for automated background

subtraction of tissue fluorescence," *J. Raman Spectrosc.* **38**(9), 1199-1205 (2007) [doi: 10.1002/jrs.1753].

[33] R. Barnes, M. Dhanoa, and S. Lister, "Standard Normal Variate Transformation and De-trending of Near-Infrared Diffuse Reflectance Spectra," *Appl. Spectrosc.* **43**(5), 772-777 (1989) [doi:10.1366/0003702894202201].

[34] K. Sowoidnich, H.-D. Kronfeldt, "Fluorescence Rejection by Shifted Excitation Raman Difference Spectroscopy at Multiple Wavelengths for the Investigation of Biological Samples," *ISRN Spectroscopy* **2012**, 256326 (2012) [doi:10.5402/2012/256326].

[35] D. G. Altman, J. M. Bland, „Diagnostic tests 1: sensitivity and specificity," *BMJ* **308** 1552 (1994) [doi: 10.1136/bmj.308.6943.1552].

[36] F. Murtagh and A. Heck, *Multivariate data analysis*, 1st ed., Springer Netherlands, Dodrecht (1987) [doi: 10.1007/978-94-009-3789-5].

[37] F. Murtagh and P. Contreras, "Algorithms for hierarchical clustering: an overview," *Wiley Interdisciplinary Reviews: Data Mining and Knowledge Discovery* **2**(1) 86-97 (2012) [doi:10.1002/widm.53].

[38] M. E. Darvin, J. W. Fluhr, P. Caspers, A. van der Pool, H. Richter, A. Patzelt, W. Sterry, J. Lademann, "In vivo distribution of carotenoids in different anatomical locations of human skin:

comparative assessment with two different Raman spectroscopy methods," *Exp. Dermatol.* **18**(12): 1060-1063 (2009) [doi: 10.1111/j.1600-0625.2009.00946.x].

[39] J. Lademann, W. Köcher, R. Yu, M. C. Meinke, B. Na Lee, S. Jung S, W. Sterry, M. E. Darvin, "Cutaneous carotenoids: the mirror of lifestyle?" *Skin Pharmacol. Physiol.* **27**(4) 201 (2014) [doi: 10.1159/000357222]

[40] A. K. Jain, R. P. W. Duin, J. Mao, "Statistical pattern recognition: a review," *IEEE Transactions on pattern analysis and machine intelligence* **22**(1) 4-37 (2000) [doi: 10.1109/34.824819].

[41] M. Gniadecka, H. C. Wulf, N. N. Mortensen, O. F. Nielsen, and D. H. Christensen, "Diagnosis of basal cell carcinoma by Raman spectroscopy," *J. Raman Spectrosc.* **28**(23), 125-129 (1997) [doi: 10.1002/(SICI)1097-4555(199702)28:2/3<125::AID-JRS65>3.0.CO;2-#].

**Elmar Schmälzlin** is a senior scientist at the Leibniz Institute for Astrophysics. He received his diploma in physical chemistry and his PhD in the field of laser spectroscopy from the Ludwig-Maximilians-Universität, Munich in 1996 and 2000, respectively. He is the author of more than 30 journal papers. His current research interests includes fiber optics, imaging Raman spectroscopy and optical sensing of oxygen.

**Ingo Gersonde** is a senior scientist at the innoFSPEC Potsdam, University of Potsdam, Germany. He received his PhD degree in physics from Freie Universität Berlin, Germany. His current work is concerned with photon density wave spectroscopy and fiber spectroscopy. His work also includes chemometrics and classification methods for analysis of spectra and sensor data.

**Martin M. Roth** is professor at the University of Potsdam, Institute of Physics and Astronomy. He is speaker of the innoFSPEC Potsdam Innovation Center and head of the innoFSPEC branch at the Leibniz Institute for Astrophysics Potsdam (AIP). He received his Diploma degree in physics and his PhD degree in astrophysics from the Ludwig-Maximilians-Universität Munich in 1986 and 1993, respectively. He is the author of more than 80 journal papers and has written two book chapters. His current research interests include resolved stellar populations in nearby galaxies, astrophotonics, and biomedical imaging.